# Science Drivers for AGN Observations with the Cherenkov Telescope Array

Henric Krawczynski

*Washington University, Physics Department and McDonnell Center for the Space Sciences*
*1 Brookings Dr., CB 1105, St. Louis, MO 63130, USA*
*Visiting Scientist at the Observatoire de Paris-Meudon in Summer 2011, 11, Avenue Marcelin Berthelot, F-92 195 Meudon cedex, France.*

*E-mail:* `krawcz@wuphys.wustl.edu`

The current generation of Imaging Atmospheric Cherenkov Telescopes (IACTs), including the H.E.S.S., MAGIC, and VERITAS telescope arrays, have made substantial contributions to our knowledge about the structure and composition of the highly relativistic jets from Active Galactic Nuclei (AGNs). In this paper, we discuss some of the outstanding scientific questions and give a qualitative overview of AGN related science topics which will be explored with the next-generation Cherenkov Telescope Array (CTA). CTA is expected to further constrain the structure and make-up of jets, and thus, to constrain models of jet formation, acceleration, and collimation. Furthermore, being the brightest well-established extragalactic sources of TeV $\gamma$-rays, AGNs can be used to probe the EBL, intergalactic magnetic fields, and the validity of the Lorentz Invariance principle at high photon energies.

# 1. Introduction

Even though state-of-the-art telescopes achieve impressive angular resolutions, they are not sufficient to spatially resolve the immediate surroundings of the supermassive black holes at the cores of AGNs. Even for the black hole in the Milky Way and the supermassive black hole in the massive elliptical galaxy M87, the VLBA radio interferometer (angular resolution: ~0.5 mas) can only resolve features with a projected angular extent exceeding ~50 Schwarzschild radii. Jets are thought to form at distances of between 10 and $10^3$ Schwarzschild radii from the black hole. So even for supermassive black holes with the largest aspect ratios, current technology only starts to be able to resolve the jet formation region. Observations in the Very High Energy (VHE) γ-ray regime (100 GeV-100 TeV) allow us to study the central engine and the jet formation process with the help of spectrally and temporally resolved data. The VHE observations constrain the make-up and structure of AGN jets, and the process of AGN accretion. At the time of writing this paper, ~45 AGNs with redshifts as high as 0.536 (3C 279) have been established as VHE emitters. In addition to blazars, VHE telescopes have detected four radio galaxies, M87, Cen A, IC310 and NGC1275. The detections of a large number of blazars can be explained by a strong observational bias: their relativistic jets are thought to be aligned to within a few degrees with the line of sight, resulting in a strong amplification of the observed luminosity by a factor of $\delta_j^4$ ($\delta_j$ being the relativistic Doppler factor). In this paper, we give a brief review of current science questions regarding the physics of AGNs (Section 2), and outline in which areas the upcoming Cherenkov Telescope Array (CTA) is expected to make major contributions (Section 3). We conclude with a summary in Section 4. The paper is limited to a qualitative discussion. The CTA collaboration is presently working on quantitative estimates based on Monte Carlo simulations of the CTA performance. The results of those studies will be presented in forthcoming papers.

# 2. AGN Science Topics

AGN topics of current interest include (i) the cosmic history of AGN assembly and growth, (ii) the inner workings of the various types of AGNs, (iii) the relation between the AGN and their host galaxies, both in terms of how the hosts switch AGNs on and off, and how the AGN feedback modifies the hosts, (iv) particle acceleration in AGNs, (v) understand the commonalities and differences between binary black hole systems (BBHs), AGNs, gamma ray bursts (GRBs), and pulsar wind nebulae (PWNs), and (vi) use AGNs as bright beacons of radiation which allow us to probe the Universe at different epochs. In the following, we briefly comment on these topics. The discussion will serve as a framework for the discussion of the CTA science drivers.

**The cosmic history of AGN assembly and growth**: We still do not know what seeds the growth of the central engines of AGNs. Observations of distant quasars indicate the presence of supermassive black holes (SMBH) with masses of a few billion solar masses less than a million years after the big bang [6, 7]. Models involving population III stars as the seeds [8, 9] face the problem that accretion seems to be too slow to explain the emergence of bright quasars at high redshifts [10, 11]. Models invoking the gravitational collapse of gas within isolated proto-galaxies may not work if star formation depletes the gas reservoir and prevents the gas from collapsing into a black hole too quickly. Direct gravitational collapse during galaxy mergers may present a feasible solution [12]. While Hubble, Spitzer, and Chandra observations have revealed new insights into the formation and growth of black holes (see e.g. [13]), substantial uncertainties remain, e.g. an understanding of the observed tight correlation of the masses of SMBHs and the properties of their host-galaxies (see [13, 14], and references therein).

**The inner workings of AGNs**: The physics of AGNs all the way from feeding AGNs to the accretion and jet formation process involve a wide range of different processes. Although recent general relativistic magnetohydrodynamic simulations have shed light on the origin of the accretion disc viscosity (the magneto-rotational instability) (e.g. [15], and references therein), the configuration and structure of an accretion system depend critically on poorly defined boundary conditions, i.e. on how the host galaxies supply gas and magnetic fields to the accretion disks [16]. Unfortunately, we know little about the long-range order of the magnetic fields in AGN accretion flows, a decisive ingredient for the formation of relativistic outflows. Detailed descriptions of the various types of AGNs and important AGN components are given in [17]. A satisfactory explanation of how some particularly simple AGN classes work, how their accretion disks are structured, how they produce jets, and what these jets are made of at different distances from the central engine, would be a major break-through.

**The relation between AGNs and their hosts**: The Chandra X-ray satellite has discovered large bubbles in galaxy clusters, devoid of thermally emitting gas (e.g. [18], and references therein). Radio observations show that the bubbles are created by non-thermal radio plasma, which displaces the hot intracluster medium (ICM). Although inflating the bubbles requires a substantial amount of mechanical *p-dV*-work, it is presently not clear if this work contributes to the heating of the ICM in a significant way. In any case, the bubbles show that AGNs interact with their environment and may have a major impact on the overall star formation efficiency in galaxies. X-ray observations with micro-calorimeter arrays with sufficiently high throughput and energy resolutions to characterize bulk motions and turbulence of the ICM are a promising avenue for advancing our understanding of the feedback process.

**Particle acceleration**: The observation of >10 TeV emission from AGNs shows that AGNs jets accelerate particles to TeV energies [19]. The outflows of AGNs may accelerate particles to much higher energies and are commonly invoked to explain the acceleration of ultra high energy cosmic rays (UHECRs), see e.g. [20], and references therein. The particles may be accelerated in shocks or in magnetic reconnection events. In the former case the central engine creates an outflow with bulk kinetic energy stored in the jet medium (electron-positron plasma or electron-ion plasma). In the latter case, the central engine creates a Poynting-flux dominated outflow. Both acceleration mechanisms involve the dissipation of macroscopic (kinetic or electromagnetic) energy to the random kinetic energy of microscopic particles. TeV observations afford excellent opportunities to study particle acceleration mechanisms. An advantage of TeV observations over longer wavelength observations is that high-energy particles (most likely electrons and positrons) radiatively cool on very short (minute) time scales, and TeV observations can track the temporal evolution of the particle energy spectra (see

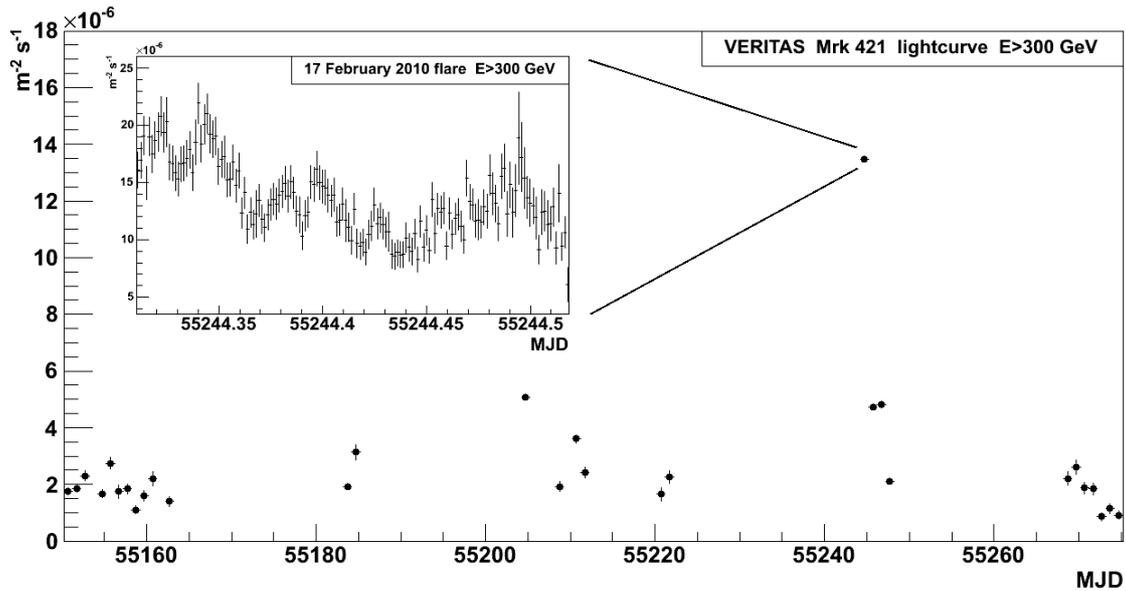

**Figure 1**: Mrk 421 fluxes measured in 2009 and 2010 with the VERITAS experiment (MJD 55160 is 11/25/2009). The small panel shows the light curve observed on 2/17/2010 with spectacular short time variability (from [2]).

Figure 1 for an example). The disadvantage of TeV observations is that most sources are spatially unresolved, rendering the unambiguous interpretation of the data difficult. In the case of shock acceleration, simulations of the non-linear interplay of collisionless shocks, non-thermal particles, and self-excited plasma waves with particle in cell (PIC) calculations present a major break-through (e.g. [21, 22], and references therein).

**Other systems with relativistic outflows**: BBHs, AGNs, GRBs, and PWNs are believed to create relativistic outflows and jets. The jets from BBHs are known to be mildly relativistic, AGN jets have bulk Lorentz factors between a few and ~50, GRB jets have bulk Lorentz factors of a few hundred, and pulsar winds are believed to be ultra-relativistic. The research aims at elucidating the differences between the accretion process in BBHs, GRBs, and AGNs, and at developing a detailed understanding of the particle acceleration processes in all four systems. Particle acceleration in ultra-relativistic shocks attracted a lot of attention (e.g. [23], and references therein). For AGNs, particle acceleration in mildly relativistic "internal" shocks within the outflow seems to be a more likely scenario than acceleration in highly relativistic "external" shocks where the jet runs into an external medium. The reason is that even a very low magnetization of the upstream jet plasma suppresses the generation of plasma waves that can serve as scattering centers [24]. Without the latter, diffusive shock acceleration is not an efficient acceleration mechanism.

**Probing intergalactic space and its contents**: AGNs have a long history as bright beacons for the study of the Hubble flow. Unfortunately, AGNs are poor standard candles, and accordingly, the most important cosmological results of the last two decades - the acceleration of the universal expansion - came from supernova observations and not from AGN observations. AGNs turn out to be the brightest extragalactic VHE γ-ray sources detected so far.

Owing to particle physics effects, the γ-ray beams can be used to address a number of long-standing astroparticle physics questions. These measurements include the measurement of the infrared/optical extragalactic background light (EBL), constraints on the strength of the extragalactic magnetic field (EGMF), and the test of Lorentz Invariance at high proton energies based on time-of-flight measurements.

## 3. CTA Science Drivers

**Determine the location of "blazar zone"**: One of the key uncertainties for the interpretation of multiwavelength observations of the blazar continuum emission is the physical location of the emission region. The observations of $\Delta t \sim 5$ min flux variability [25-30] limit the size of the emission region to $R_\gamma < \delta_j \Delta t\, c \sim 7.2 \times 10^{14}\, (\delta_j/40)\, (\Delta t/10\, \text{min})$ cm. Theories involving particle acceleration in the jet usually place the emission region between 30 and a few thousand $R_{\text{Sch}}$ from the black hole, with $R_{\text{Sch}}$ the Schwarzschild radius ($R_{\text{Sch}} = 3 \times 10^{14}$ cm for a $10^9\, M_\odot$ black hole). The observations of exceptionally strong γ-ray flares from M87 in 2008 observed by H.E.S.S., MAGIC and VERITAS in temporal coincidence with an exceptionally strong VLBA radio flare from the M87 radio core were interpreted as evidence for an origin of the γ-ray flares within a projected distance of 50 $R_{\text{Sch}}$ from the central engine [31]. Unfortunately, a similar γ-flare in 2011 was not accompanied by a comparable radio flare and did not corroborate the association. CTA will afford the possibility to perform VLBA/VHE correlation studies with greatly improved VHE sensitivity. The study would require several years of regular observations of a sample of key-sources, e.g. M87, BL Lac, Mrk 501 and 1ES 1959+650 with CTA and the VLBA. The recent discovery of a strong flare from the Flat Spectrum Radio Quasar (FSRQ) PKS 1222+21 ($z=0.43$) with the MAGIC IACT system was used to set a lower limit on the distance $d_\gamma$ between the VHE emission region and the central engine of $d_\gamma > 3 \times 10^{17}$ cm [32]. A smaller distance would result in the absorption of the detected VHE γ-rays by photons from the broad line region (BLR) clouds. Interestingly, the MAGIC data show evidence for a flare on a time scale of 10 min, setting an upper limit on the size of the emission region of $R_\gamma < 1.8 \times 10^{14}\, (\delta_j/10)\, (\Delta t/10\, \text{min})$ cm. Combining the lower limit on $d_\gamma$ with the upper limit on $R_\gamma$, the authors infer that the emission region subtends an angle $\theta_j < \arctan(R_\gamma/d_\gamma) = 0.03°$, much smaller than the likely opening angle of the jet of a few degrees inferred from source statistic arguments [33]. CTA is expected to improve substantially on these results. CTA will discover VHE emission from a statistical sample of FSRQs and will set stringent limits on $R_\gamma$ for the brightest sources. The combination of CTA observations with optical spectroscopy observations before, during, and after CTA observations will enable the derivation of robust constraints on $d_\gamma$.

**Jet-structure and jet-composition at the base of the jet**: Over the last 20 years, various γ-ray instruments (EGRET, Whipple, HEGRA, CAT, H.E.S.S., MAGIC, VERITAS, Fermi and others) have been used together with radio, optical, and X-ray telescopes to gather simultaneous broadband energy spectra of blazars. The modeling of the available spectral energy distributions (SEDs) with leptonic synchrotron-Compton codes can be summarized as follows (e.g. [34-36], and references therein). (i) One-zone synchrotron self-Compton (SSC) models - in which a single population of electrons emits synchrotron emission, and inverse

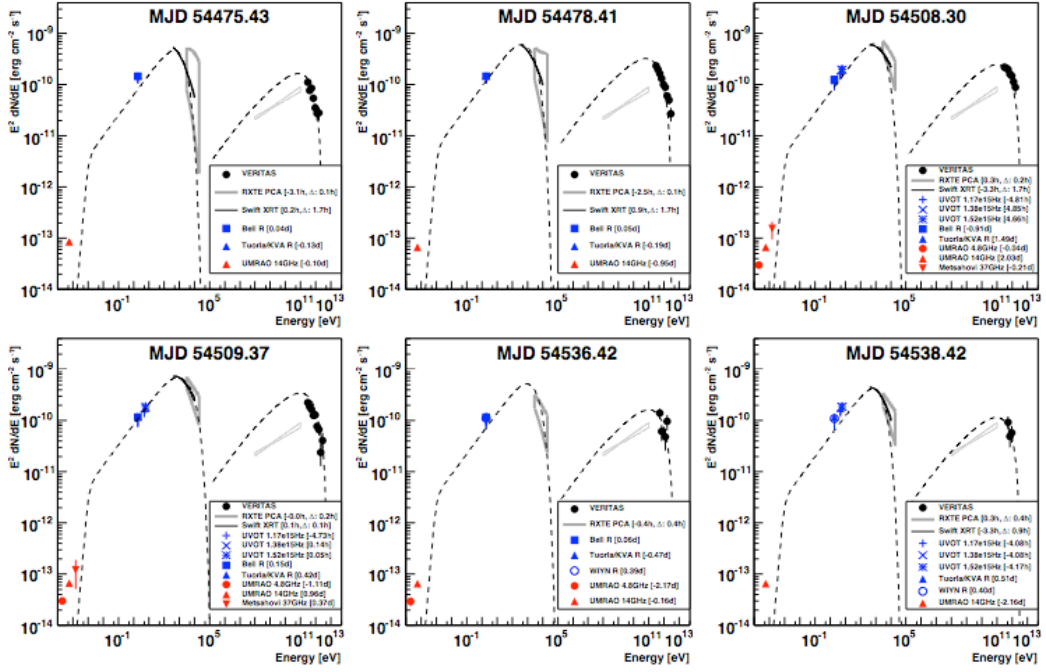

**Figure 2**: One-Zone Synchrotron Self-Compton models of Mrk 421 spectral energy distribution for a set of multiwavelength observations (from [1]).

Compton emission from interaction of the electrons with the synchrotron photons - give a satisfactory description of most data sets; for some data sets, the SEDs of multiple flares can be described with the same model, changing only one or two of the 8 model parameters of the simplest SSC models (*electron energy spectrum:* normalization, low & high-energy cutoff, break energy, spectral index; *emission region:* radius, magnetic field, Doppler factor); see Figure 2 for exemplary SSC fits; for other data sets, fitting multiple flares requires an adjustment of several model parameters or the introduction of multiple electron components or emission zones (e.g. [37, 38]). (ii) Although some data sets can be fitted with low Doppler factors on the order of $\delta_j \sim 5$, many SEDs require Doppler factors of $\delta_j \sim 50$ (e.g. [36]); for a few sources, observations of rapid flares on time scales of ~2 min also indicate Doppler factors of the order of 50 [25, 26]. (iii) SSC models favor particle-energy-dominated emission regions with low magnetizations $\sigma = B^2/8\pi / \int d\gamma \, (dN_e/d\gamma) \, \gamma \, m_e c^2 < 1/100$ (e.g. [1, 34]). If additional external target photons are invoked, larger emission volumes with lower electron densities per unit volume can fit the data, and can raise the inferred magnetization values to $\sigma \sim 1$ [39].

For FSRQs, Sikora & Madejski (2000) argue that the jet cannot be dominated by cold pairs at the jet base, as the inverse Compton emission from the pairs scattering cosmic microwave background (CMB) photons would give rise to an unobserved soft X-ray emission component [40].

CTA is expected to improve our knowledge of the jet-structure and jet-composition in the blazar zone in several ways. CTA will measure blazar energy spectra over a broader energy range with exquisite accuracy (<< 0.1 in spectral index). Simultaneous X-ray/VHE coverage (if available) would allow us to search not only for X-ray/VHE flux correlations, but also for X-ray/VHE spectral index correlations. These measurements would make it possible to test SSC models even in the case that multiple parameters change during flares. Simultaneous optical and soft X-ray/VHE observations of FSRQs will allow us to improve on the reliability of the limits on the cold pair content of jets.

**Jet structure and composition at kpc-distances from the central engine**: The Chandra satellite revealed X-ray emission from kpc-scale quasar jets as a common phenomenon [41]. The X-ray emission could be synchrotron emission from TeV electrons; alternatively it could be inverse Compton emission from rather cold electrons scattering photons from the CMB if the jet plasma moves with relativistic bulk Lorentz factors even at kpc-distances from the central engine. VHE observations have the potential to distinguish between these two scenarios as the TeV electrons in the former scenario would up-scatter the CMB and produce an observable VHE component. The emission from the kpc-scale jet can be distinguished from emission from the core based on the lack of time variability, the spectral slope of the spectrum which is predicted to be the same as in the X-ray band in the inverse Compton CMB scenario, and the spatial offset of the kpc jet emission from the position of the core. A VHE detection would not only vindicate the synchrotron model for the X-ray emission, but would also allow us to measure the bulk Lorentz factor of the jet medium at kpc-distances from the central engine. Stawarz et al. describe the possibility to discover hard VHE emission from X-ray-bright jets with FR I radio morphology [42]. For these sources, the synchrotron spectrum seems to extend all the way from the radio to the X-ray regime. Stawarz et al. argue that the same population of TeV electrons which produces the observed X-rays as synchrotron emission should emit VHE γ-rays from SSC processes and inverse Compton scatterings of ambient and CMB photons. The authors predict the CTA detection of steady or slowly variable emission from Cen A and M87, and emphasize the importance of such detections for constraining the jet bulk Lorentz factor and magnetic field.

**Particle acceleration in jets**: X-ray observations of blazars with Chandra, XMM-Newton, Suzaku, and Swift give energy spectra with excellent signal to noise ratios – even for relatively short integration times – and can show us today what CTA will do in the future. Of course, the real benefit will come from simultaneous broadband observations of blazars, which emphasizes the importance for a rapid deployment of CTA to take advantage of X-ray and γ-ray coverage afforded by the current and/or upcoming X-ray and γ-ray missions. Figure 3 shows light curves and hardness ratio from an observation of the blazar Mrk 421 from May 5 to May 9, 2008 with the Suzaku satellite [3]. Suzaku can constrain the spectral index to an accuracy of ~0.02 for 4000 s observation windows. One of the interesting features is a slow hardening of the energy spectrum from $t = 2\times10^4$ s to $10^5$ s after the start of the observations when the flux decayed slowly. Later, around $t = 2\times10^5$ s, a similar flux decay is accompanied by a softening of the spectrum. Although the spectrum hardens *during* the flare at $t = 1.6 \times10^5$ s into the observations, it hardens *after* the flare at $t = 2.6 \times10^5$ s. The data show that each flare behaves differently. The authors argue that the observations imply a temporal change of the spectral index and/or the high-energy cut-off produced by the acceleration mechanism. Explaining the high-energy cut-off in the framework of particle acceleration at shocks as a consequence of the competition between acceleration gains and synchrotron energy losses, sets a lower limit on the factor $\eta$, the ratio between the mean free path and the gyro-radius, of $\eta > \sim 10^5$. Such high ratios seem unlikely, as they would imply inefficient injection of the electrons into the acceleration mechanism.

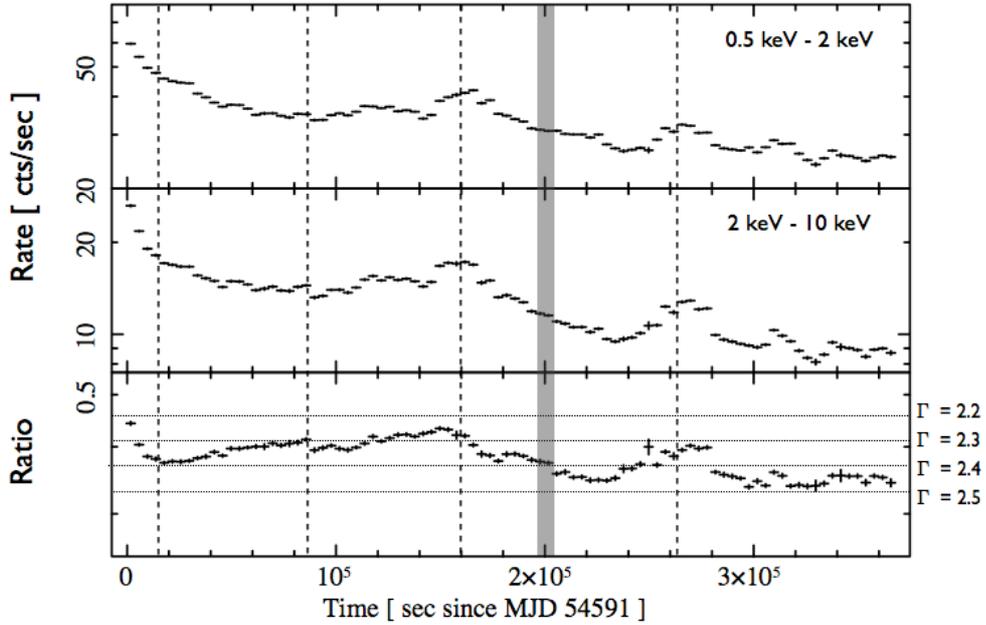

**Figure 3**: Suzaku Mrk 421 light curves and hardness ratios ((2-10 keV counts) / (0.5-2 keV counts)) with 4000 s binning. Observations were taken from UTC 2:52am, May 5, 2008 to UTC 8:24am, May 9, 2008. The horizontal dotted lines in the lower panel give the hardness ratios for power-law spectra of several power-law photon indices $\Gamma$ (from [3]).

      The authors propose a stronger magnetic field in the particle acceleration region than in the emission region as a solution to this problem. It is interesting to compare the results from the SED modeling of blazars (see the paragraph "Jet-structure and jet-composition at the base of the jet" above) with the predictions of the theories of particle acceleration at shocks and in magnetic reconnection events. Electron acceleration seems to be more efficient in weakly magnetized shocks ($\sigma \ll 1$) [21, 22]. Magnetic reconnection predicts an emitting plasma with a magnetization of $\sigma > \sim 1$ (e.g. [43, 44]). The SSC modeling gives σ < 1/100, clearly favoring shock acceleration over reconnection as the particle acceleration mechanism. If the emission is of leptonic origin, a pair-dominated plasma seems to be favored over an electron-ion plasma, as the latter preferentially accelerates ions – and electron energy spectra tend to be soft. If the emission is of hadronic origin, the plasma should obviously be an electron-ion plasma.

      The interpretation of combined X-ray/CTA data with a quality comparable to that shown in Figure 3, would benefit from the additional constraints on the model parameters afforded by the complementary X-ray/VHE information. Time dependent SSC models (see Figures 2-4 of [34]) predict spectral index changes of ~ 0.02 during flares. In view of such small changes, it is clear that the control of systematic errors on spectral indices will be of utmost importance for CTA. Atmospheric monitoring accompanying all observations will be important to extract the scientific results allowed by the excellent photon statistics.

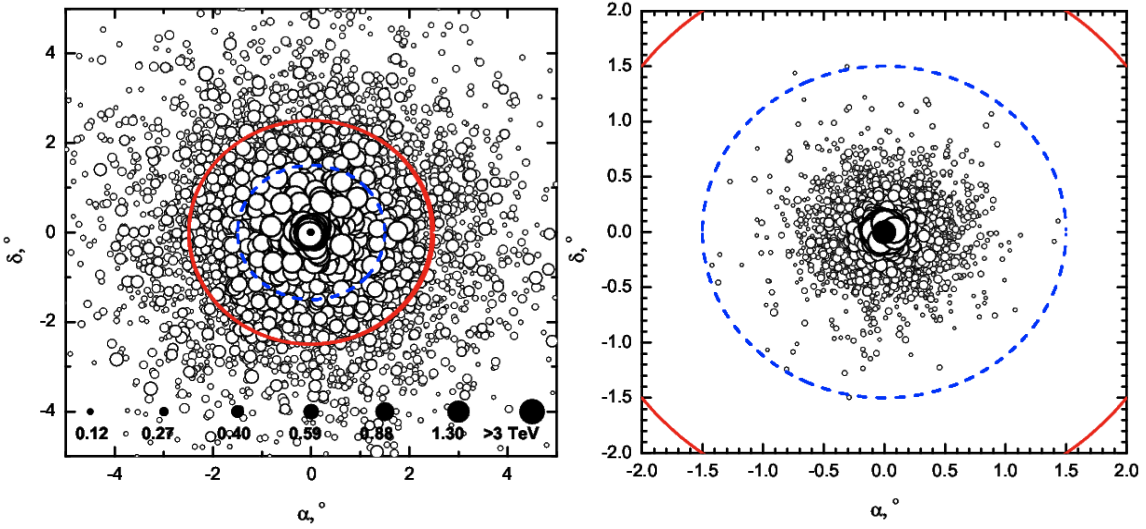

**Figure 4**: Arrival directions of primary and secondary cascade γ-rays (circles) from a source at a distance D = 120 Mpc for an EGMF of $10^{-14}$ G (left panel) and $10^{-15}$ G (right panel). Each photon is shown as a circle with a diameter depending on its energy. The blue (red) circle has a radius of 1.5˚ (2.5˚). The figure has been reproduced with the kind permission of the authors of [5].

**Constraints from Pair Cascades and Pair Haloes**: On their way from the AGNs to us, a fraction of the VHE photons are absorbed owing to $\gamma_{VHE} + \gamma_{IR/O} \rightarrow e^+ e^-$ pair-production processes on infrared/optical EBL photons. As the EBL is made of direct and reprocessed emission from stars and AGNs at all redshifts, and possibly also from decays of exotic particles and population III stars in the early Universe, the measurement of its intensity and energy spectrum are of great interest. Although extragalactic pair-production processes reduce the visibility of high-redshift objects, they open up exciting possibilities to measure the intensity and energy spectrum of the IR/optical EBL, and to constrain the strength of the mean extragalactic magnetic field (EGMF). Combined H.E.S.S., MAGIC, VERITAS and Fermi data have already been used to set the first observational lower limits on the strength of the EGMF (see [45], and references therein). The following chain of arguments is used. The EBL attenuates the γ-ray energy spectra of the considered sources (RGB J0710+591 at $z$=0.13, 1ES 0229+200 at $z$=0.14, and 1ES 1218+304 at $z$=0.18). Some of the pairs created in the extragalactic pair creation processes inverse Compton scatter CMB photons into the MeV, GeV or TeV energy bands. The combination of pair creation and inverse Compton processes gives rise to an extragalactic electromagnetic cascade. Some of the energy emitted at TeV energies will show up in the Fermi energy range as cascade emission. However, for a vanishing intergalactic magnetic field, the detected VHE fluxes already imply MeV-GeV fluxes in excess of the ones observed by Fermi. The only solution to this apparent contradiction is an EGMF with $B > 10^{-17}$-$10^{-15}$ G (assuming the magnetic field is uniform on spatial scales of 1 Mpc). The EGMF will deflect the cascade electrons and lead to a spatial and temporal spread of the cascade emission. The dilution of the emission in angular space and/or in time reduces the expected cascade fluxes to a level consistent with the Fermi observations. The limits on the EGMF derived with this procedure are very interesting for cosmologists and particle physicists.

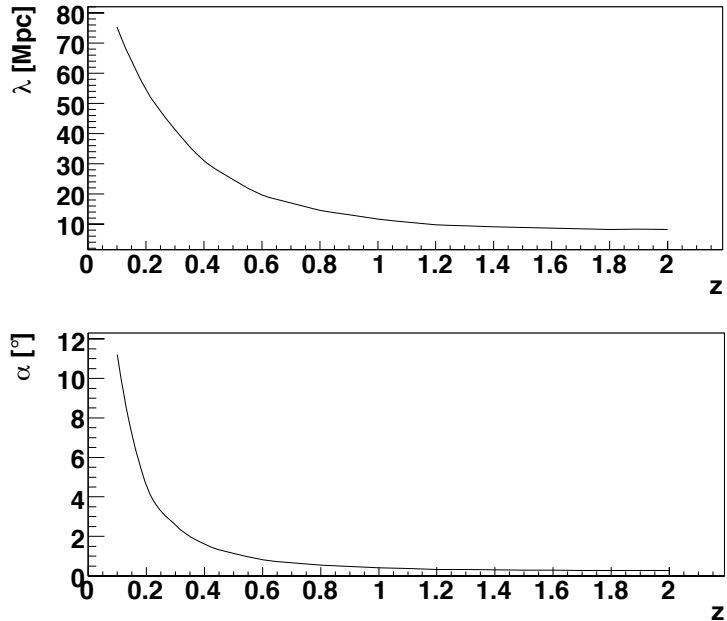

**Figure 5**: Radius $R_H$ in Mpc (top panel) and angular radius in degrees (bottom panel) at which the γ-ray emissivity of a pair halo drops by a factor of 1/e as function of redshift. The calculation assumes 10 TeV primary photons so that the inverse Compton photons from the first generation of $e^+$ and $e^-$ have an energy of ~100 GeV in the observers' frame. The calculation uses the optical/infrared background from [4], and assumes a strong EGMF which isotropizes the electrons on spatial scales smaller than $R_H$.

The EGMF may have been created in phase transitions in the early Universe, or by stars, galaxies, and galaxy clusters.

CTA will elevate this type of analysis to the next step by enabling a uniquely sensitive search for the cascade emission. The angular and temporal properties of the cascade emission would give a measurement of the EGMF (see e.g. [5, 45-51] and references therein). If the EGMF is close to the present lower limits, the detection of the cascade emission from the *z*=0.1-0.2 blazars should be possible. High signal-to-noise measurements would allow us to compare the energy-dependent radial profile of the halo emission with cascade simulations for different magnetic fields (Figure 4).

If the EGMF is strong enough to isotropize the cascade electrons, but not strong enough to lead to rapid synchrotron cooling of the secondary particles which quenches the development of an inverse Compton/pair creation cascade, "pair haloes" from distant AGNs may be detectable [46, 52]. Figure 5 shows the expected sizes of pair haloes as a function of the redshift for haloes generated by 10 TeV γ-rays whose first generation inverse Compton γ-rays have an energy of ~100 GeV. The calculation uses the background model of Franceschini et al. 2008 [4] and assumes a strong EGMF which isotropizes the electrons on spatial scales much smaller than the halo diameter. Although the haloes are very large for redshifts *z* < 0.5, a halo with a radius <1˚ at *z* > 0.5 would fit into the field of view of CTA telescopes and would allow for reliable background subtraction. If one assumes that the source has a constant $\nu F_\nu$-flux from 100 GeV to 10 TeV, a simple estimate shows that the halo-fluxes are below a couple of percent compared to the 100 GeV emission that reaches us directly. The halo fluxes would be much larger if the photon index at the source was much harder than $\Gamma=2$.

**Lorentz invariance tests and EBL studies**: CTA will allow us to probe Lorentz invariance based on photon time-of-flight measurements. Such tests are complementary to tests based on polarimetry, and to searches for threshold effects [53-56]. Assuming a modified dispersion relation holds: $v/c = 1 + \eta_1 E_\gamma / E_{Pl}$ with $E_\gamma$ the photon energy and $E_{Pl}$ the Planck energy, current experiments constrain $\eta_1$ to be smaller than ~1. Compared to present experiments, and assuming a source spectrum with photon index $\Gamma=3$, CTA could improve on these limits by a factor of ~5 because of constraining the time of flight over a larger energy span, and by a factor of ~2 by detecting fast flux variability of more distant blazars. If *GRBs* (rather than AGNs) are detected at GeV/TeV energies, existing limits could be improved by more than four orders of magnitude through the detection of flux variability on millisecond time scales.

TeV γ-ray observations of extragalactic objects have been used to constrain the intensity of the optical/infrared background radiation (e.g. [57-59]). Its intensity and energy spectrum are important cosmological observables for theories of galaxy and star formation. Most authors constrain the optical/infrared background radiation based on the assumption that the emitted blazar energy spectra cannot have photon indices $\Gamma$ (from $dN_\gamma /dE \propto E^{-\Gamma}$) harder than $\Gamma = 1.5$. However, several authors pointed out that the emitted energy spectra may well be harder. Possible reasons include multiple target photon components giving rise to spectral bumps, contributions from different leptonic and hadronic emission components, particle acceleration at relativistic shocks which can give electron energy spectra harder than two if large angle scattering plays a non-negligible role [60, 61], and γ-ray absorption inside the source [61, 62]. With the detection of a large number of blazars, statistical treatments start to be possible. The detection of average trends in the energy spectra should be treated with caution as such trends could be caused by selection biases. An unambiguous detection of the effect of extragalactic extinction may be possible if a pronounced absorption feature is detected in several sources at an energy that varies with redshift in the expected way.

**Discovery of new source classes**: Recently, Swift discovered the transient J164449.3+573451 at a redshift of $z=0.354$ [63]. Strong X-ray flares were detected over a time period of three days followed by a slow decay of the X-ray brightness. Based on fast flux variability and the apparent super-Eddington brightness of the source, the authors conclude that the emission is relativistically boosted, and posit that the flares were caused by the tidal disruption of a star by a dormant AGN. CTA would be ideally suited to search for VHE emission from such an event – if alerted by a suitable transient detector, e.g. Swift or the proposed JANUS and Lobster IR/GRB observatories. CTA might also be able to detect extended VHE emission from the Fermi bubbles or from radio galaxies like Cen A.

**4. Summary and Discussion**

VHE observations sample the non-thermal emission from AGNs. Although VHE emission requires extremely violent processes and particle acceleration to >TeV energies, a large number of AGNs have already been detected in the VHE regime. VHE observatories, especially the proposed CTA observatory, can scrutinize the extremely rapid time variability of the VHE fluxes and energy spectra. The rapid variability and high luminosities indicate that the flares originate close to the SMBHs, and VHE astronomy thus affords the opportunity to study relativistic outflows close to where they form. Compared to Fermi, CTA will have a much larger collection area ($10^6$ m$^2$ compared to <1 m$^2$) and a much better typical angular resolution. The large collection area makes it possible to scrutinize flares on minute time-scales with good photon statistics. The good angular resolution will make it possible to derive arc-second source localizations and to bring extended sources – like Cen A – into a sharper focus. On the flip side, CTA will have a much smaller field of view than Fermi and a lower duty cycle (~10% compared to ~100%). Thus, CTA will likely not be competitive with Fermi in terms of constraining AGN luminosity functions and their cosmic evolution. The fact that CTA will sample AGNs at the high-energy end of the electromagnetic spectrum leads to exciting prospects – i.e. the detection of intergalactic cascades, which carry information about the EBL and the EGMF. Besides AGNs, CTA will be able to detect other extragalactic sources, normal galaxies (like M31), starburst galaxies and ultra-luminous infrared galaxies, and maybe also galaxy clusters.

**Acknowledgements**

HK thanks S. Thibadeau J. Krawczynski for proofreading the manuscript. He thanks the Observatoire de Paris-Meudon for a summer research fellowship, and is grateful to H. Sol, A. Zech, and C. Boisson for interesting discussions during the visit. HK acknowledges NASA for support from the APRA program under the grant NNX10AJ56G, the DOE for support from its high-energy physics division, and support from the McDonnell Center for the Space Sciences at Washington University.